\documentclass[lettersize,journal]{IEEEtran}
\usepackage{amsmath,amsfonts}
\usepackage{algorithmic}
\usepackage{algorithm}
\usepackage{array}
\usepackage[caption=false,font=normalsize,labelfont=sf,textfont=sf]{subfig}
\usepackage{textcomp}
\usepackage{stfloats}
\usepackage{url}
\usepackage{verbatim}
\usepackage{graphicx}
\usepackage{cite}
\usepackage[colorlinks,
linkcolor=blue,       
anchorcolor=blue,  
citecolor=blue,       
urlcolor=blue,       
]{hyperref}

\begin{document}

\title{Analysis on Blockchain Consensus Mechanism Based on Proof of Work and Proof of Stake}

\author{\IEEEauthorblockN{1\textsuperscript{st} Shi Yan}
	\IEEEauthorblockA{ \\
		\textit{Wuhan University}\\
		Wuhan, China\\
		2020302181047@whu.edu.cn}
}

\markboth{Journal of \LaTeX\ Class Files,~Vol.~14, No.~8, August~2021}%
{Shell \MakeLowercase{\textit{et al.}}: A Sample Article Using IEEEtran.cls for IEEE Journals}

\IEEEpubid{~}

\maketitle

\begin{abstract}
In the white book of Bitcion, Satoshi Nakamoto described a bitcoin system that can realize point-to-point online payment without a third-party organization. After supporting this magical application scenario and subverting the traditional centralized system, the blockchain technology has attracted worldwide attention, triggered a research upsurge of blockchain consensus algorithm, and produced a large number of innovative applications. Although various consensus algorithms continue to evolve with the iteration of blockchain products and applications, Proof of Work (POW) and Proof of Skake (POS) algorithms are still the core of consensus algorithms. This paper discussestwo algorithms of POW and POS in blockchain consensus mechanism, and analyzes the advantages and the existing problems of the two consensus mechanisms. Since consensus mechanism is the main focus of blockchain technology and has many influencing factors, this paper discusses the current problems and some improved ideas, and selectes some typical algorithms for a more systematic introduction. In addition, some important issues related to safety and performance are also discussed. This paper provides the researchers a great reference on blockchain consensus mechanism.
\end{abstract}

\begin{IEEEkeywords}
Blockchain, Proof of Work, Proof of Stake, Nothing-at-stake Attack, Consensus Mechanism.
\end{IEEEkeywords}

\section{Introduction}
\IEEEPARstart{S}{ince} the emergence of bitcoin \cite{1,2} system in 2009, its underlying blockchain technology has gradually attracted the attention of academia and industry, and has achieved rapid development. At present, blockchain technology has been widely used.

In the traditional network, the reliability of the online payment platform, such as WeChat and Alipay, is determined by the provider of the business system. The security of personal identity information and transaction records also depends on the reputation and security of the third party organizations. This centralized business processing mode has been broken in the blockchain system. With the brief description of bitcoin principle in Nakamoto's study \cite{1} on November 11, 2008, as well as the official launch of bitcoin system and the advent of Genesis block in 2009, the bitcoin system has been developing tepidly for nearly 10 years. After that, people find that through the underlying technology of blockchain, it can realize the same function in a completely decentralized environment, compared to the traditional centralized environment. The key to complete this function is the consensus algorithm.

The core of the consensus mechanism is to solve the problem of consistency in the distributed system through the consensus algorithm, so that each node can be recognized and cannot be tampered with. In the blockchain system, it mainly solves the consistency of transaction data and transaction status of each node in the distributed system, and realizes decentralized multi-party mutual trust. Consensus algorithm is the core technology of blockchain and the key technology to realize the Internet from information interconnection to value interconnection.

Blockchain can be divided into public chain and license chain according to node permissions. Public chain means that blockchain nodes have no access mechanism and participating nodes can join or exit at any time. There are two representative consensus algorithms.One is proof of work (POW), which is the probability of obtaining rights by consuming computing power. The greater the computing power is, the higher the probability can be. Its advantages are the strong randomness and great fairness, while its disadvantages are energy consumption and low consensus efficiency. The second is the proof of stake (POS), which is the probability of obtaining rights by holding assets. The more assets are, the greater the probability can be. Its advantage is energy conservation and environmental protection, while the disadvantage is power concentration.

As for the consensus mechanism which is core of blockchain technology, this paper aims to discuss the POW and POS mechanisms in the blockchain consensus, and the advantages and the existing problems of the two consensus mechanisms based on the relevant studies.

\section{Different Types of Consensus Mechanisms}
\subsection{Proof of Work}
\subsubsection{Hashcash: inspiration for proof of work}
\par~ \par Proof of work (POW) for Bitcoin blockchain borrows the idea of Hashcash, which was proposed to deal with the issues of junk email in May, 1997 \cite{3}. In junk email, an attacker almost costs nothing. Software of junk email can send emails in batches to the destination address. Hashcash can solve these problems by making attackers consume some computing resources.

To be detailed, the messages of email sent by the sender must contain a signature consisting of the recipient's address, sending time, and counter. The counter is supposed to make the signature meet the conditions. The SHA-1 algorithm should be used to generate a 160 bits-long hash value for the signature, the first 20 bits of which are all zero. As a result, before sending a message, the sender needs to constantly adjust the counter value to generate a qualified email signature, while the mail server only needs to perform SHA-1 calculation once to judge whether the signature meets the conditions.

\subsubsection{Concept of POW}
\par~ \par The concept of POW was first proposed by Jakobsson in 1999 \cite{4}. POW is used to achieve verifiable computing tasks. It includes two sides: certifier and verifier. The certifier provides evidence to the verifier to show that it has completed a number of calculations in a period of time. Since the generation of evidence requires amount of computing resources, POW can be used to solve denial of service (DOS) attacks.

In bitcoin network \cite{1}, given a value $D$, there is a block contains all transaction data. Miners should calculate the ‘nonce’ so that the value calculated by SHA-256 twice is less than D, as presented as Formula \eqref{eq1}:
\begin{equation}\label{eq1}
SHA256(SHA256(data||nonce))\le D
\end{equation}

\begin{figure}[htpb]
	\centering
	\includegraphics[width=0.85\linewidth]{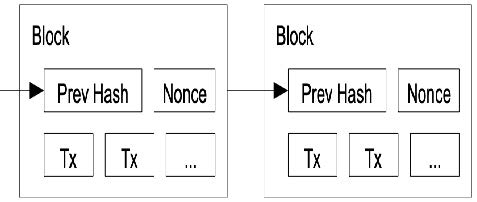}
	\caption{Nonce in blockchain node \cite{1}}
	\label{fig:01}
\end{figure}

In that case, the miner can win the competition only by constantly changing the input nonce value \cite{5}. Figure \ref{fig:01} shows nonce in blockchain node. It is determined by the computing power of the miner, so the competition of POW finally turns to the competition of hash computing power in bitcoin network \cite{5}. In addition, the value D can change to adjust the difficulty of mining \cite{1}. 

The adjustment of difficulty value occurs automatically in each complete node. Every 2016 blocks, all nodes will adjust the difficulty value according to the unified formula. This formula is obtained by comparing the spending time of the latest 2016 blocks with the expected time. The expected time is 20160 minutes, which is the total time calculated according to the generation rate of a block in every 10 minutes. According to the ratio of the actual time to the expected time, the difficulty will be adjusted.

Compared with Hashcash, bitcoin uses the SHA-256 algorithm. In bitcoin network, the first 32 bits of hash value should be zeros, and then the bitcoin network will periodically reset the difficulty to face with the increasing computing power of miners.

\subsubsection{Application of POW in Bitcoin}
\par~ \par In the bitcoin white paper, Satoshi Nakamoto proposed the application of POW in bitcoin network \cite{1}. He described the following six points:
\begin{itemize}
	\item New transactions are broadcast to all nodes. 
	\item Each node collects new transactions into a block. 
	\item Each node works on finding a difficult proof-of-work for its block. 
	\item When a node finds a proof-of-work, it broadcasts the block to all nodes. 
	\item Nodes accept the block only if all transactions in it are valid and not already spent. 
	\item Nodes express their acceptance of the block by working on creating the next block in the chain, using the hash of the accepted block as the previous hash.
\end{itemize}

\begin{figure}[htpb]
	\centering
	\includegraphics[width=0.72\linewidth]{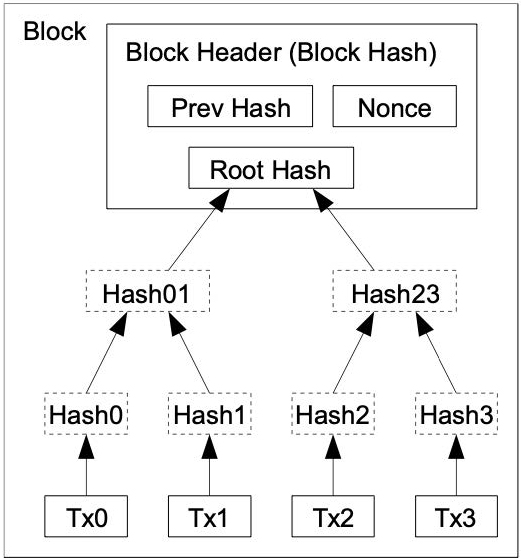}
	\caption{Transactions hashed in a Merkle Tree \cite{1}}
	\label{fig:02}
\end{figure}

In bitcoin network, to realize the process of POW, we should first generate Merkle root hash through Merkle tree algorithm. The figure \ref{fig:02} shows transactions hashed in a Merkle Tree. After that, we judge whether 2016 blocks have been generated after the last difficulty value adjustment. If the condition is satisfied, the algorithm will adjust the difficulty. If the condition is not satisfied, we take the 80B data of the current block header, prev-block hash, bits, nonce, Merkle root, and timestamp as the input values of the POW. Furthermore, we continuously adjust the nonce, and perform double SHA-256 operation on the block header after each adjustment. Finally we compare the calculated value with the current difficulty value $D$. If it is less than the difficulty value, the mining is successful.

\subsubsection{Three Issues of POW}
\paragraph{The Centralization of Computing Power}
\par~ \par The proof of work(POW) of bitcoin is computationally intensive, which is easy to lead to the centralization of network computing power. In the bitcoin white paper \cite{1}, Satoshi Nakamoto proposed the concept of "one CPU one vote". In Nakamoto's vision, nodes can use personal computers to perform POW operations, participate in block node elections, and receive corresponding compensation. However, with the rise of bitcoin price, the block reward has attracted a large number of computing power to join, and the hash computing power in bitcoin network shows an exponential growth trend \cite{6}. The physical equipment of the consensus node participating in POW computing has been transformed from the early personal computer to GPU, and then evolved into the ASIC miner widely used at present.

In view of the computationally intensive characteristics of SHA-256 hash function, some blockchain systems choose to replace the original function with memory intensive hash function. For example, Litecoin \cite{7} and Dogecoin \cite{8} adopt the script algorithm \cite{9}, Ethereum \cite{10} adopts the Ethash algorithm \cite{11}, and Zerocash \cite{12} and Zerocoin \cite{13} adopt the Equihash algorithm \cite{14}. Memory intensive hash function can reduce the computing power advantage of ASIC mining machine to a certain extent because it occupies a lot of memory and is difficult to compute in parallel.

\paragraph{Waste of Resources}
\par~ \par POW of Bitcoin will lead to the waste of computing resources. The existing literature \cite{15} estimates that the annual power consumption of bitcoin network is equivalent to that of Ireland or Austria.

In order to solve the problem of resource waste, some blockchain systems use the computing power consumed in POW computing to provide useful services. For example, primecoin \cite{16} improved the pow problem to find prime numbers that meet the requirements, thus promoting the development of the field of mathematics.

\paragraph{Performance}
\par~ \par Since the average block interval of bitcoin system is 10min and the block size is limited to 1MB, theoretically, the transaction throughput is about 7 transactions per second. Low throughput limits the wide application of bitcoin system. With the increasing attention of bitcoin system, the number of unconfirmed transactions in the network increases. As of July 10, 2019, there were nearly 50000 unconfirmed transactions on bitcoin network \cite{17}. The performance problem has become an urgent problem to be solved in bitcoin pow.

In view of the low performance of bitcoin pow, some research work and blockchain system improve efficiency by modifying parameters and improving block node election mechanism. For example, Ethereum, Wright coin and dogcoin systems adjust the block interval in the bitcoin POW mechanism to 15s, 2.5min and 1min \cite{18} respectively, so as to accelerate the transaction processing speed. Shortening block spacing seems to be a feasible scheme to improve performance. However, some research work has found that shortening block interval has potential security risks.

\subsection{Proof of Stake}
\subsubsection{Proposal of POS}
\par~ \par POW consensus algorithm submits a result which is difficult to calculate but easy to verify through the compute power competition among all nodes. Every node can verify the result through simple operation, so as to recognize the work of the verifier. This mechanism effectively maintains the security of bitcoin system. However, the unfair computing power, waste of resources, and low efficiency of POW algorithm seriously limit the expansion of bitcoin blockchain technology.

In view of the waste of resources in the POW, the bitcoin community \cite{19} proposed proof of stake (POS) in 2011 for the first time. The algorithm stipulates that the number of bitcoins owned by nodes is used to replace the process of solving hash value based on computing power in POW. The more bitcoins owned by miners, the greater the possibility of mining. The security of the POS mechanism is based on the fact that equity owners are more motivated than miners to maintain network security \cite{1}. If the blockchain system is attacked, the interests of equity owners will be more likely to be damaged. In 2012, the proof of stake mechanism was applied in peercoin system for the first time \cite{20}.

\subsubsection{Proposal and Concept of Coin-days}
\par~ \par In the application of blockchain, it is impossible for us to truly allocate shares to nodes in the chain. Instead, other things are used as shares, and we allocate these things to nodes in the chain. 

In the application of POS, the amount of currencies can be recorded as the number of equity. Now Ethereum has the POS consensus mechanism. Therefore, in Ethereum, the number of shares of each Ethereum node is measured by the number of Ethers. For example, in an Ethereum network with three nodes, A, B and C, A has 100 Ethers, B and C have 10 and 20 respectively, and then A block is the most likely one to be selected.

However, If only based on the coins number, it will lead to the result of centralization, because the richest members will have permanent advantage.

Peercoin's proof of stake system combines the concepts of random selection and Coin-Days. Coin-Days is the number of currencies multiplied by the time they are held. When Coins held for more than 30 days, user will have the opportunity to become the producer of the next block. The users with larger coin-days will have a greater chance to sign the next block. However, once a block has been signed, its Coin-days will be cleared to zero. The user must wait another 30 days to obtain the chance to sign the next block. The coin-days of currency will only add up to 90 days and will not increase, in order to avoid users of very large coin-days having an absolute role in the block chain.

\subsubsection{Concept of POS}
\par~ \par Given a difficulty value D of the whole network and the trade data newly packaged into the block, the qualified counter can be found, as presented as Formula \eqref{eq2}:
\begin{equation}\label{eq2}
SHA256(SHA256(tradeDate||Counter))<D\times Coindays 
\end{equation}

The accounting rules of the POS consensus algorithm are basically similar to the POW consensus algorithm, while the POS consensus algorithm does not require miners to enumerate all the random numbers nonce, but only allows one hash value within 1s, which greatly reduces the computational work and curb the resource consumption in computing power competition.

\subsubsection{Application of POS}
\par~ \par Peercoin takes the Coin-ages as the equity to elect block nodes. The concept of Coin-ages is also applied in Cloakcoin and Novacoin \cite{21}. It can suppose that user A has 10 coins and holds them for 90 days, with a cumulative Coin-ages of 900. User B has 10 coins and holds them for 45 days, with a cumulative Coin-days of 450. According to concept of POS, user A is twice as likely to solve the problem as user B.

In 2014, Pavel Vasin proposed POS2.0 \cite{22} and applied it to blackcoin, so that the Black Coin developed from POW + POS consensus mode to pure POS consensus mode at that time. The rights equity in the POS2.0 consensus algorithm are directly proportional to the user's online time. This incentive mechanism effectively enhances the blockchain P2P network, prevents the occurrence of "tragedy of the commons" in the pow consensus algorithm, and greatly increases the difficulty for attackers to realize "51\% attacks". However, POS is prone to bifurcation, and the decentralization of blockchain is weakened.

\subsubsection{Nothing-at-stake Attack}
\par~ \par Nothing at stake attack refers to the problem that the selected out of block node generates multiple blocks at the same height, resulting in block fork. As shown in Figure 3, the blockchain system based on the proof of stake forks at height $h$, and node A is the block out node at height $H + 1$. Since the block finally confirmed at height $h$ is not determined, node A selects to generate blocks after the multiple forked blocks at the same time. Therefore, no matter which block is legal in the end, node A can make a profit.
\begin{figure}[htpb]
	\centering
	\includegraphics[width=0.95\linewidth]{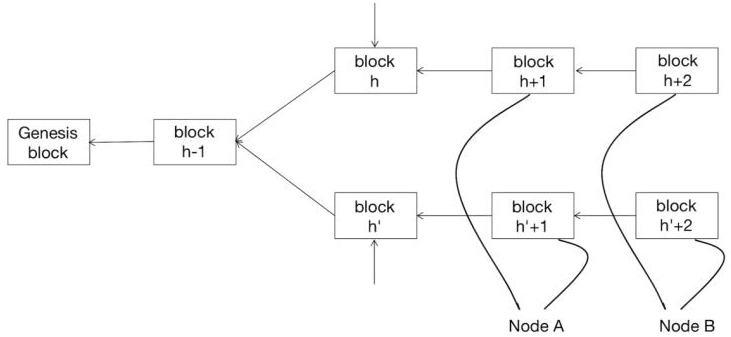}
	\caption{Nothing-at-stake attack}
	\label{fig:03}
\end{figure}

The attack cost of nothing-at-stake attack is low, and the out of block node can generate multiple blocks without any cost. Therefore, the multiple forked blocks go hand in hand and never reach a consensus. The existing solutions to non equity attacks mainly include margin system and security hardware. Margin system means that the consensus node needs to deposit and withdraw a certain amount of margin in the account. If the node is monitored to launch a nothing-at-stake attack, its margin will be confiscated to solve the nothing-at-stake attack from the perspective of economic incentive. For example, both slasher and Casper \cite{23} of Ethereum's POS proposal have introduced the margin system. Similarly, Tendermint \cite{24} introduced a margin mechanism.

\section{Conclusion}
Blockchain technology is accelerating the development of digital economy in the process of integration with real industries. As the core technology of blockchain, the consensus algorithm can achieve the consistency of results on a specific transaction in a network environment where nodes are highly dispersed and there is no mutual trust mechanism, and there is no disagreement on the process and results. Generally, the choice of consensus algorithm is to balance efficiency, security and stability on the premise of specific application scenarios.

This paper focuses on the two aspects of POW and POS in the blockchain consensus mechanism, and analyzes the advantages and the existing problems of the two consensus mechanisms. Some typical algorithms are discussed in detail in this paper. Based on the current research status, in view of the development of blockchain consensus algorithm, it is necessary to study and solve the problems of performance and security. This paper is a great references for the researches studying on blockchain consensus.

%
\section{Simple References}
You can manually copy in the resultant .bbl file and set second argument of $\backslash${\tt{begin}} to the number of references
 (used to reserve space for the reference number labels box).

\vfill
\end{document}